\documentclass[sigconf]{acmart}

\AtBeginDocument{%
  \providecommand\BibTeX{{%
    \normalfont B\kern-0.5em{\scshape i\kern-0.25em b}\kern-0.8em\TeX}}}

\usepackage{adjustbox}
\usepackage{array} % for centering text in cells
\usepackage{booktabs} % For better-looking tables
\usepackage{subcaption}
\usepackage{graphicx}
\usepackage{multicol}
\usepackage[toc,page]{appendix}
\usepackage{multirow}

\newcommand{\sysname}{Savant}
\newcommand{\syscomp}{AccessBolt}
\newcommand{\command}{NLC}
\newcommand{\controlvalue}{<CE,~Value(CE)>}
\newcommand{\ignore}[1]{}

\newcommand{\fixmevk}[1]{{\bf\textcolor{orange}{ [ vk FIXME: #1 ]}}}

\author{Satwik Ram Kodandaram}
\authornote{Both authors contributed equally to this research.}
\email{skodandaram@cs.stonybrook.edu}
\affiliation{%
  \institution{Stony Brook University}
  \city{Stony Brook}
  \state{New York}
  \country{USA}
}
\orcid{0000-0002-5208-4301}

\author{Utku Uckun}
\authornotemark[1] % This links back to the first author note
\authornote{This research was done while the author was at Stony Brook University.}
\email{uuckun@cs.stonybrook.edu}
\affiliation{%
  \institution{Stony Brook University}
  \city{Stony Brook}
  \state{New York}
  \country{USA}
}
\orcid{0009-0007-1580-7308}

\author{Xiaojun Bi}
\email{xiaojun@cs.stonybrook.edu}
\affiliation{%
  \institution{Stony Brook University}
  \city{Stony Brook}
  \state{New York}
  \country{USA}
}
\orcid{0000-0002-9716-7709}

\author{IV Ramakrishnan}
\email{ram@cs.stonybrook.edu}
\affiliation{%
  \institution{Stony Brook University}
  \city{Stony Brook}
  \state{New York}
  \country{USA}
}
\orcid{0000-0002-1768-7043}

\author{Vikas Ashok}
\email{vganjigu@odu.edu}
\affiliation{%
  \institution{Old Dominion University}
  \city{Norfolk}
  \state{Virginia}
  \country{USA}
}
\orcid{0000-0002-4772-1265}


\begin{document}

\title{Enabling Uniform Computer Interaction  Experience for Blind Users through Large Language Models}

\begin{abstract}

Blind individuals, who by necessity depend on screen readers to interact with computers, face considerable challenges in navigating the diverse and complex graphical user interfaces of different computer applications. The heterogeneity of various application interfaces often requires blind users to remember different keyboard combinations and navigation methods to use each application effectively. To alleviate this significant interaction burden imposed by heterogeneous application interfaces, we present~\sysname{}, a novel assistive technology powered by large language models (LLMs) that allows blind screen reader users to interact uniformly with any application interface through natural language. Novelly, \sysname{} can automate a series of tedious screen reader actions on the control elements of the application when prompted by a natural language command from the user. These commands can be flexible in the sense that the user is not strictly required to specify the exact names of the control elements in the command.  
%The design of~\sysname{} was informed by the findings of a Wizard-of-Oz study with $11$ blind computer users. \fixmeiv{This study was not used in the esign.}
A user study evaluation of~\sysname{} with $11$ blind participants demonstrated significant improvements in interaction efficiency and usability compared to current practices.

\end{abstract}


\maketitle

\section{Introduction}

% The ubiquitous use of computing applications in educational, professional, and personal contexts and the development of assistive technologies have substantially improved the employment prospects of individuals with visual impairments~\cite{bell2015employment, wang2010making}. Numerous employment opportunities of this nature necessitate proficiency with desktop productivity tools such as the Microsoft Office suite~\cite{bovenzi2003enabling}. The Graphical User Interface (GUI) of these desktop applications is mainly designed for two-dimensional interaction, relying on visual cues, such as buttons, menus, and icons, to enable effortless \textit{``point-and-click''} interaction with a computer \textit{mouse} or \textit{touchpad}. To interact with these applications, blind users primarily rely on screen reader technology such as JAWS \footnote{\url{https://www.freedomscientific.com/products/software/jaws/}}, NVDA \footnote{\url{https://www.nvaccess.org/}}, and VoiceOver \footnote{\url{https://www.apple.com/accessibility/mac/vision/}}. A screen reader functions in a linear, one-dimensional fashion by reading and interpreting content displayed on the screen sequentially. For instance, \textit{File}, \textit{Edit}, \textit{View}, and other \textit{Ribbon} options via a limited keyboard shortcut vocabulary (e.g., \textit{Alt+Arrow} to move between options in dropdown menus). 

People who are blind depend on assistive technology, specifically a screen reader (e.g., JAWS~\footnote{\url{https://www.freedomscientific.com/products/software/jaws/}}, VoiceOver~\footnote{\url{https://www.apple.com/accessibility/mac/vision/}}, and NVDA~\footnote{\url{https://www.nvaccess.org/}}), to interact with computer applications. The screen reader reads the content on the screen, allowing blind users to navigate and interact with different control elements and content sections of the application using keyboard shortcuts. However, modern computer applications feature complex and densely packed graphical user interfaces (GUIs) that are mainly designed for visual \textit{point-and-click} interactions using a mouse, rather than relying on keyboard shortcuts, the principal method of interaction for blind individuals. As a result, blind people often face difficulties in performing basic tasks (such as accessing ribbon commands) in applications~\cite{lee2020repurposing}.
 
These interaction challenges are compounded by the heterogeneity of the applications' interfaces that require blind users to remember specific keyboard shortcuts and navigation methods for each of these applications.  For example, in VLC Media Player, the keyboard shortcut to play or pause audio is ``Spacebar'', while in Windows Media Player, it is ``Ctrl + P''. Although both applications fulfill the same purpose, blind users must recall and distinguish between these unique keyboard shortcuts, increasing their cognitive burden and possibly impeding their interaction efficiency. In addition, different screen readers often require different keyboard shortcuts to execute identical tasks, resulting in differences in user interaction between platforms. For instance, in \textit{ NVDA}, the standard key combination to access a menu is usually \textit{NVDA + N}, while in JAWS, it is \textit{Insert + F10}. These variations in shortcuts persist when performing the same task in versions of identical applications on various platforms, e.g., in Word on Windows, the key combination Alt + Ctrl + F is used to add a footnote, while in the MacOS version of Word it is Command + Option + F.
These challenges have arguably hindered blind users in effectively mastering even accessible desktop applications, with research indicating that they are frequently ten times slower than sighted individuals when performing similar computer tasks~\cite{buzzi2012designing,hae-na,uckun2022taming}.  

\begin{figure*}[!t]
    \centering
    \includegraphics[width=15cm,height=7.5cm]{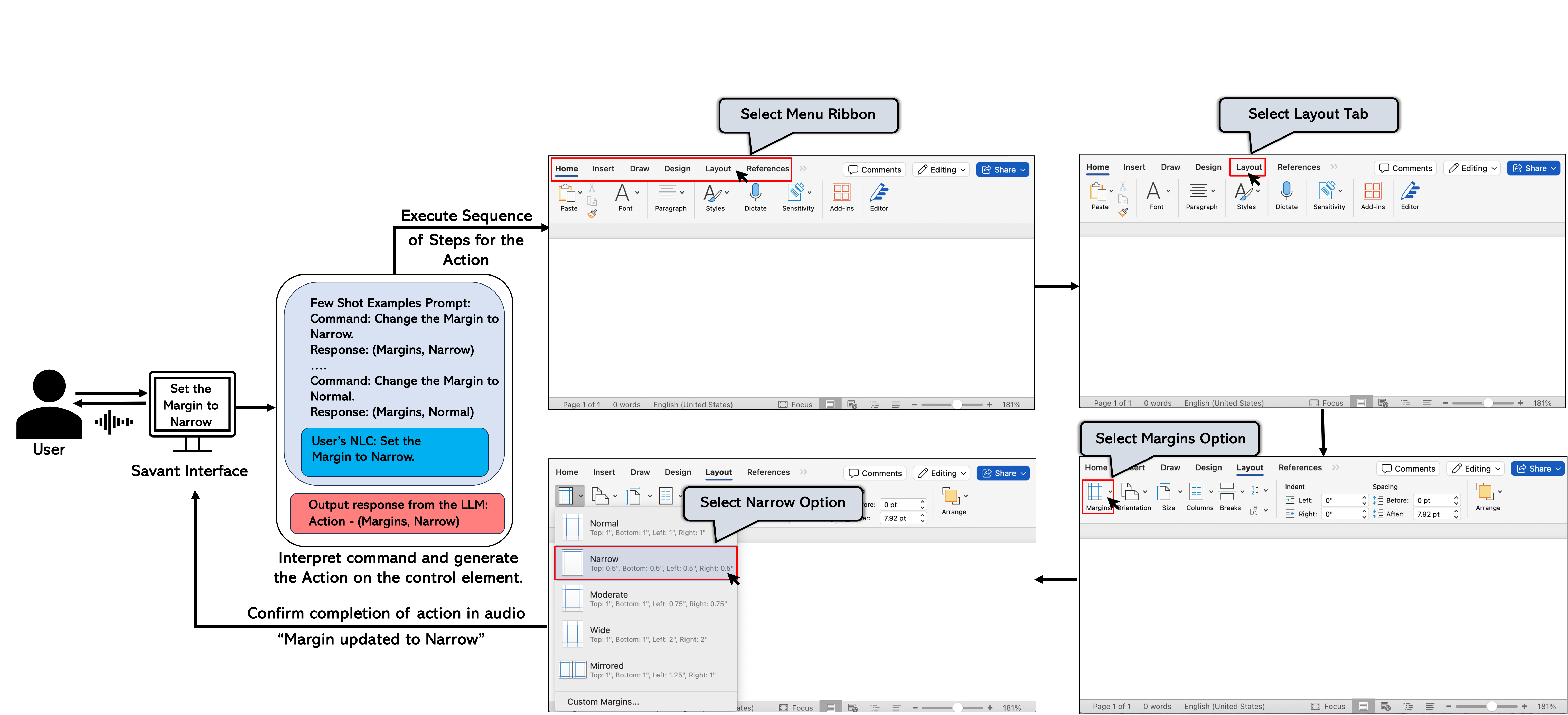}
    \caption{LLM-powered automation of sequence of screen reader steps prompted by the user’s command in natural language}
    \label{fig:teaser}
\end{figure*}

In a previous study, blind participants emphasized the importance of having uniform and consistent access independent of the specific application, screen readers, and platforms~\cite{billah2017ubiquitous}, and the absence of this type of access in contemporary assistive technology was considered a significant drawback by all study participants. Assistive technology that incorporates approaches to offer this type of uniform access sought by blind users has not received much attention and is therefore the focus of our work in this paper.

We specifically introduce~\sysname{}, a novel assistive technology powered by large language models (LLMs) that augments the screen reader for enabling blind users to interact uniformly with any application interface using natural language. Uniquely, \sysname{} simplifies and speeds up interactions with screen readers by automating a sequence of laborious screen reader actions on the application's control elements when triggered by a natural language command (\command{}) from the user. These commands offer flexibility as they do not strictly require the user to identify the precise names of the control elements in the \command{}.

As an illustration of \sysname{}, consider Joe, a blind individual preparing a Word document. Suppose, for starters, that he wants to alter the standard margin ``normal'' to ``narrow''. 
Should Joe were to rely solely on his screen reader, his first step would be to engage the \textit{Menu Ribbon} by hitting \textit{ALT + H}. Following this, he would traverse the tab list using the \textit{right arrow} key to shift focus to the \textit{Layout} tab. Subsequently, he would explore the \textit{Layout} choices using the \textit{Down Arrow} key, opt for \textit{Margins}, and finally select \textit{Narrow}. 
On the other hand, with \sysname{}, Joe simply needs to say ``set the margin to narrow'' to do the same task, and in response, \sysname{} with the aid of an LLM, interprets the intent in Joe's transcribed command as an Action on a
control element. It then identifies the series of screen reader steps required to implement the Action and autonomously executes this sequence of steps for Joe.   

As depicted in Figure~\ref{fig:teaser}, \sysname{} correctly interprets the intention of the spoken command ``set the margin to narrow'' as an Action on the pair <Margins,\ Narrow>, which amounts to assigning the value ``Narrow'' to the control element ``Margins''. 
It identifies the series of screen reader steps required to implement the Action <Margins,\ Narrow>, namely, \textit{Select Menu Ribbon,  Select Layout Tab, Select Margins, Select Narrow Option}, to implement the margin change, and finally executes them on the Word application interface in the correct sequential order. Note that in the absence of \sysname{}, these tedious steps would have to be done manually by Joe.
Significantly, even if Joe had said ``change the margin to narrow'', or ``update the margin to 0.5'', or ``I want to create a narrow margin'', \sysname{} would still have accurately interpreted all of these commands given the powerful reasoning capabilities of its LLM, leading to the same series of screen reader actions on the Word application interface, underscoring the power and flexibility of \sysname{}.
Finally, \sysname{} enables Joe to use the same command to adjust margins in other similar applications (e.g., WordPad, Pages) as well, thereby obviating the need to remember distinct shortcuts for these applications. Thus, \sysname{} offers a promising approach to achieving consistent access across diverse interfaces.

LLMs belong to a class of foundational models that are trained on vast amounts of data, enabling them to comprehend, analyze, and produce natural language and various other forms of content for a wide variety of tasks. Their notable successes and impact on numerous fields have gained them widespread recognition. 
However, leveraging their capabilities in assistive technologies, as exemplified by the \sysname{} system outlined in this paper, is a relatively unexplored topic.

\sysname{} harnesses the capabilities of LLMs in text comprehension to offer a natural language command interface for uniform interaction across various application interfaces. The design of \sysname{} is described in Section~\ref{savant:design}. To assess the effectiveness of \sysname{} in practice, we conducted a user study with $11$ blind participants. \sysname{} showed significant improvements in both perceived usability ($3$ times higher SEQ score) and task workload ($3$ times lower NASA-TLX score) over status-quo screen readers, for typical computer tasks that involve accessing controls in a diverse set of applications. Participants were also able to access application controls nearly $4$ times faster on average than with screen readers alone while doing these tasks. The accuracy of LLM-based interpretation and automation of participant commands was also high with a $0.78$ accuracy score.

In summary, the contributions of this paper are the design, development, and evaluation of a novel assistive computer technology for blind users, namely \sysname{}, which takes advantage of LLM to provide uniform and efficient natural language-based access to the controls of any computer application. It should be noted that Savant could also benefit the elderly or those with motor impairments who may struggle with the precision required for traditional mouse and keyboard input.

\section{Related Work}

\subsection{Computer Usability for Blind Users}

% People with severe visual disabilities including blindness primarily use a screen reader to interact with computer applications. A screen reader is an assistive technology software that narrates application content and enables users to navigate the content using dedicated shortcuts (e.g., TAB for next GUI element).  

% The popularity of screen readers among the blind user's community is because of the fact it is relatively cheaper and easy to use compared to hardware devices such as braille displays~\cite{haga2005dynamic, xu2011tactile, yobas2003novel}. Screen Readers offer a variety of shortcuts that enable visually impaired users to navigate the GUI elements of an application in a one-dimensional paradigm and facilitate efficient navigation and content accessibility ~\cite{nvdashortcuts, jawsshortcuts, voiceovershortcuts}. 

% Nevertheless, the abundance of shortcuts provides a significant challenge for visually impaired users since it may be arduous to remember and effectively employ each shortcut. In general, individuals tend to depend on a limited number of fundamental shortcuts in order to navigate computer applications~\cite{borodin2010more}.

While there have been plenty of works addressing the accessibility of computer applications~\cite{singh2012blind, doush2013non, zou2015chartmaster, islam2023probabilistic, kodandaram2023detecting, sunkara2023assessing}, including the availability of guidelines and recommendations~\cite{morales2013design, harper2012web}, the research on usability, i.e., the ease, efficiency, and satisfaction with which blind users can do their computer tasks, is still in its infancy, with very few efforts investigating the usability of select computer applications~\cite{wentz2011usability, leporini2012interacting, ashok2018non}. As modern computer applications all have two-dimensional graphical user interfaces primarily designed for sighted interaction, there is an inherent mismatch with the one-dimensional navigation supported by a screen reader, thereby resulting in numerous usability issues as uncovered by the findings of prior studies~\cite{baldwin2017tangible, uckun2022taming, wentz2011usability, miao2016contrasting, wentz2013survey}. For instance, an early work by Shinohara et al.~\cite{shinohara2007observing} illuminated the navigational difficulties faced by blind users in computer applications. Their study notably revealed that even a single accidental key press could utterly disorient a blind user in the application UI, and moreover, recovery from such errors often entailed significant mental effort and time. In another study, Billah et al.~\cite{billah2017ubiquitous} observed that differences in shortcut mappings between different screen readers and between different applications generated confusion and interaction problems for blind screen reader users. A recent work by Islam et al.~\cite{islam2023probabilistic} revealed that screen reader users' perceptions of usability and accessibility of desktop applications are nuanced, incorporating various factors beyond the traditional definitions. These factors include independence, ease of learning an application, ability to describe it to others, reliability and deterministic behavior of keyboard shortcuts, and transfer ability of knowledge to others.

% \fixmevk{Discuss papers only related to usability issues.}

Apart from the above works studying the usability issues in computer applications, there also exist research works proposing assistive-technology solutions for improving the usability of computer applications. However, these works have largely centered on the web ~\cite{borodin2010more, sunkara2023enabling, ashok2023assistive, ferdous2022insupport, ferdous2023enabling} and mobile applications~\cite{ leporini2012interacting, ko2021modeling, li2023modeling, li2022select, billah2019accessible}. For instance, quite a few natural language assistants and web automation techniques have been proposed in the literature for enabling convenient web interaction for blind users~\cite{ashok2015capti, ashok2017web}. Assistive technologies based on third-party customized hardware have also been presented to improve navigational efficiency in web screen reading. Similarly, there have been other works that have tried to improve usability by modifying the DOM of the web pages to better suit screen-reader navigation~\cite{aydin2020sail}. 

% \fixmevk{These are so old and irrelevant. Can add recent ones??}
% \fixmevk{find some new ones!!}

In contrast to these web and mobile usability solutions, assistive technologies for enhancing desktop-application usability have been relatively scarce~\cite{ morales2013design, ashok2023assistive}. Moreover, almost all these extant works have primarily targeted certain specific aspects of select computer applications. For instance, a few works in this regard~\cite{lee2020rotate, buzzi2012designing, mori2011making, schoeberlein2014usability, waqar2019intelligent} have specifically targeted usability of certain features of word-processing applications, such as ribbon controls, collaborative elements such as user comments, and dynamic document updates. Similarly, Doush et al. \cite{doush2013non} created a multi-modal solution for customizing non-visual navigation and reading of tabular data in the \textit{Microsoft Excel} application. There also have been application-specific usability solutions for PDF software~\cite{uckun2020breaking, uckun2020ontology}, Email application~\cite{wentz2013survey}, and Office applications~\cite{Office-Applications}. On the other hand, Uckun et al.~\cite{uckun2022taming} proposed a general system-wide solution for the Windows OS platform, namely \textit{AccessBolt}, that leveraged Microsoft's UI Automation tree to provide an alternative popup interface for sequentially accessing application controls \textit{in-one-place}, instead of navigating all over the application GUI to access these controls. While AccessBolt showed significant improvements in usability in a user study with blind participants, it also exhibited certain key limitations. First, AccessBolt could only provide access to `visible' controls on the screen; controls not in view (e.g., controls in dropdown menus, inactive ribbons) were not accessible in AccessBolt's popup interface. Moreover, sequential navigation of controls in the popup still requires considerable time and shortcut presses, especially in complex applications such as productivity tools that have hundreds of controls. In contrast, \sysname{} shields the user from all these low-level screen-reader operations and related GUI-navigation intricacies, thereby enabling a more convenient and uniform interaction via flexible natural language commands for blind screen-reader users.

\subsection{Natural Language Assistance for Blind Users}
Many commercial voice-activated assistants are now available on desktops and smartphones, such as Apple's Siri~\cite{Siri}, Google Assistant~\cite{Google-Assistant}, and Microsoft's Cortana~\cite{Cortana}, that enable users to automate certain computer tasks such as setting reminders and alarms, taking notes, making Facebook/Twitter updates, etc. However, these assistants provide little-to-no support for in-application activities such as accessing and invoking controls~\cite{ashok2018non, ashok2015capti}. For example, a user can ask Siri to open the Word application on MacOS but cannot instruct it to \textit{set the margins}, \textit{track changes}, etc. Even in native MacOS applications such as Finder, Siri is not able to automate simple tasks such as renaming a file

% ~\fixmevk{Verify this. It did not work on my Mac. Check if windows Cortana can do in-application tasks -- Cortana can't do in-application tasks}.

In addition to the aforementioned system-wide assistants, there also exist application-specific assistants (typically offered by the application vendor) that enable users to access features and automate tasks within the corresponding applications. For example, Google Docs has an in-built assistant that can automate document formatting activities such as font, size, and color~\footnote{\url{https://support.google.com/docs/answer/4492226?hl=en&ref_topic=1361462&sjid=13413164066125249767-NA}}. Recently, Amazon has enabled a natural language assistant namely Rufus~\footnote{\url{https://www.aboutamazon.com/news/retail/amazon-rufus}} in their mobile application for select customers, to enable quick and easy access of product-related information via natural language queries. Similarly, many other apps are providing their own chatbots and assistants to enhance their usability~\cite{sidlauskiene2023ai}. The obvious drawbacks of these assistants are that they are tightly bound to their host application and moreover the underlying technologies are closed source, therefore, it is difficult if not impossible to replicate these assistants for general use across all applications.  

Other than the above assistants in commercial software products, there also exist research prototypes that offer greater flexibility in terms of the range of supported commands and are also more generic in that they support multiple applications~\cite{ashok2018non, gadde2014screen}. For instance, Ashok et al. proposed natural language interfaces for blind screen reader users to conveniently and efficiently navigate web applications using a wide range of spoken natural language commands~\cite{ashok2015capti}. Specifically, these assistants first translate the user's spoken commands into the intended browsing actions and then automatically execute these actions on the user's behalf. JustSpeak~\cite{zhong2014justspeak} is another assistant service on the Android mobile platform that can also map user utterances to basic actions such as scroll, toggle switch, toggle checkbox, and activate controls. A common aspect of these aforementioned research prototypes is that they target either web or mobile applications. Phutane et al.~\cite{phutane2023speaking} explored the potential of conversational assistants to enhance non-visual access for blind and low vision individuals via human-like conversations. To the best of out knowledge there is no existing application-agnostic, natural language-based assistive technology for desktop applications that allows for consistent and efficient interaction across various applications and platforms. Extant prototypes for web and mobile cannot be readily adapted for desktop usage due to the inherent heterogeneity in the underlying software implementations and technologies. \sysname{} is envisioned to fill this gap.

\section{Technical Design of \sysname{}}
\label{savant:design}
\begin{figure*}
    \centering
    \includegraphics[width=15cm,height=8.5cm]{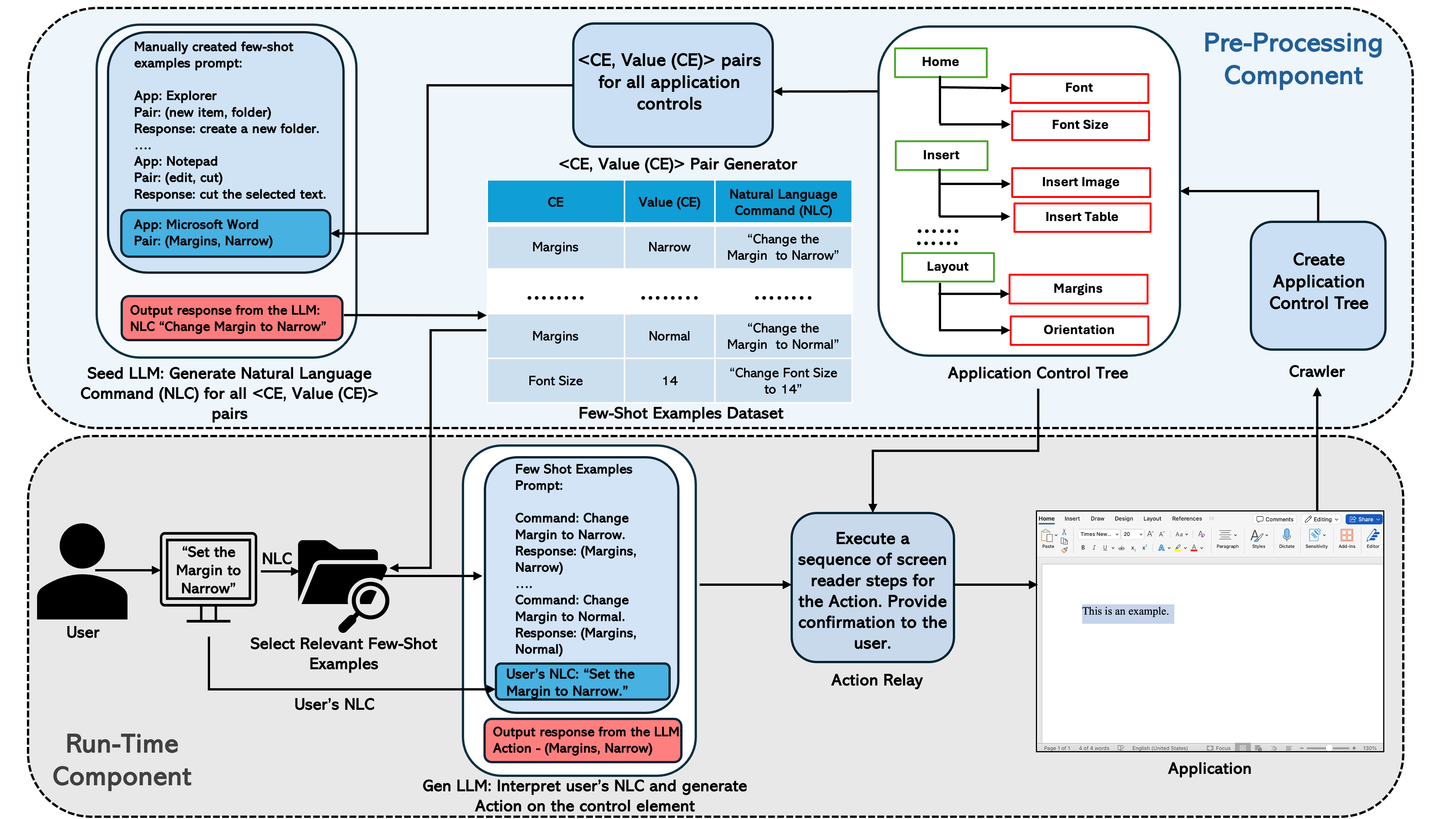}
    \caption{Architectural workflow of the~\sysname{} system. CE stands for Control Element. Value (CE) stands for the value of the Control Element. NLC stands for Natural Language Command.}
    \label{fig:technical-diagram}
\end{figure*}

{\bf Notations:} 

%We will use \textit{Action} to denote a user action such as clicking, selecting, etc., on the interface elements.

We use \textit{NLC} to denote a natural language command.

We will use \textit{CE} to denote control elements. A CE is an interface element on which the user can execute click, select, etc. 

We will use \textit{Value(CE)} to denote a value the CE can take. 

We associate an  \textit{Action}  with the pair <CE, Value(CE)>, meaning execution of the Action will result in a sequence of screen reader steps that will result in the assignment of Value (CE) to CE.
We will use CE, control elements, and controls interchangeably.

\subsection{Preliminary Data Collection}
\label{woz}
To understand the types of \textit{NLCs} that \sysname{} can expect to receive from blind users while they interact with computer applications, we conducted a Wizard-of-Oz (WOZ) study with $11$ blind participants. In the study, the participants were asked to do an assortment of computer tasks in $5$ different applications including \textit{Excel}, \textit{Gmail}, \textit{File Explorer}, \textit{Word}, and \textit{Zoom}. The full details of the study including the design, tasks, participant details, and procedure, are provided in Appendix~\ref{subap:wizard-of-oz-study}. Overall, from the WOZ study, we collected a total $145$ NLCs, and each NLC was manually annotated with the \controlvalue{} pair of the corresponding application. This dataset provided the important \textit{seed} data for both our model training and testing as explained next. 

\subsection{Overview}

\label{overview}

\sysname{} system architecture consists of two components, a Preprocessing component, and a Runtime component -- see Figure~\ref{fig:technical-diagram}. In the figure, the upper half, shown in blue, denotes the Preprocessing component, whereas the Runtime component is shown in gray.

The objective of the Preprocessing piece is two-fold: (1) to create a non-visual representation of the application  GUI that will contain information that will be used to generate the sequence of screen reader steps to invoke an action on a control element (see ``Application Control Tree''  in the figure); (2)  to develop a dataset of few-shot samples for the application to serve as prompts (shown as ``Few-Shot Examples Dataset'' in the figure).  

Note that preprocessing is done \textit{only once} for each application. The application's Application Control Tree (ACT) and its Few-Shot Examples Dataset (FED) are repeatedly used at runtime to generate the sequence of screen reader steps that are required to execute the user's \command{} as shown in the Runtime Component in Figure~\ref{fig:technical-diagram}.
%Note that FED is a collection of few short examples for all applications of interest to the user.

The \textit{Crawler} module in the figure constructs the ACT by employing the UI Automation tool~\cite{UIAutomation-webpage} to extract all the control elements of the application, inclusive of meta-data (e.g., control element names, control types, hierarchical relationships between control elements, actions on the control elements like clicking, selection, etc.). 

The \textit{Pair-Generator} module is responsible for the creation and storage of \controlvalue{} pairs for every control element in the application, using the ACT's meta-data as a reference. To illustrate, consider Microsoft Word. 
Since one possible value for  \textit{Margins} is \textit{Narrow}, the resulting pair would be \textit{(Marin, Narrow)}. Similarly, it can take other possible values such as Normal, Moderate, Wide, etc., and a corresponding pair for each such value will be created.

The \textit{Seed LLM} module generates the few-shot examples dataset for the application and saves it in the FED. Each example in the FED is a triple <NLC, CE, Value(CE)>. The semantics of the triple provide an interpretation for \command{}, that is, linking the action of assigning Value(CE) to the control element CE.

For this purpose, we provide prompts to the \textit{Seed LLM} to generate the triples in FED. The prompt for \textit{Seed LLM} includes a few hand-crafted few-shot examples and a~\controlvalue{} pair originating from the \textit{Pair-Generator} module. The LLM output is an NLC corresponding to the~\controlvalue{} pair and is stored in the FED. The hand-crafted prompts instruct the Seed LLM on the creation of the NLC for each~\controlvalue{} pair. It is important to note that these hand-crafted prompts are done once. They are generic and are used in the generation of the FED dataset for any application.

The objective of the Runtime component is to interpret the user's \command{}, generate the correct \controlvalue{} pair based on the interpretation, associate an Action on the \controlvalue{} pair, and execute a sequence of screen reader steps corresponding to the Action.

Gen LLM carries out the interpretation and mapping of the NLC to the correct \controlvalue{} pair. The prompt it uses comprises the user's NLC input along with a collection of few-shot examples that are pertinent to the NLC. The \textit{Action Relay} module generates the sequence of steps in the screen reader to execute the Action on the  \controlvalue{} pair. To this end, it references the information in ACT that facilitates the execution of the step sequence. Details of the two components are provided in the following sections.

\subsection{Preprocessing Component}

Recall that the Preprocessing Component will build the ACT ( the Application Control Tree)  and the FED ( the few-hot examples dataset). Their construction is described in the following sections. 

\subsubsection{\textbf{Constructing Application Control Tree}}

We developed a custom \textit{Crawler} module to build the ACT made up of all the control elements of the application. In the first step, the \textit{Crawler} module utilizes the Microsoft UI Automation tool~\cite{UIAutomation-webpage} to extract all visible controls of the application and other related metadata. Visible controls are those seen on the screen of the application window (e.g., Control Name, Control Type, etc.).
However, controls that are not visible (e.g., sub-menus) typically do not appear in the UI Automation tree unless a user interaction prompts them. Extracting them is a bit tricky. The approach we use is as follows.

To extract these hidden controls, the \textit{Crawler} explores visible control elements breadth-first and automatically simulates user interactions on each of these elements, such as clicking on drop-down menus or tabs, to uncover hidden controls embedded within the visible controls.
For example, in Microsoft Word, when the ``font family'' drop-down is expanded, it reveals all the selectable font types. When the ``Insert'' tab is selected, it shows all the controls in that tab.  By executing these simulated interactions, invisible controls are made visible in the UI Automation tree.
The \textit{Crawler}  captures these newly revealed controls along with their metadata as mentioned in the overview in Section~\ref{overview}. This process is done recursively for nested controls (e.g., sub-menus). Throughout this recursive process, the \textit{Crawler} maintains a record of the actions that reveal new controls and associates these actions with the corresponding control elements, thus generating a hierarchical tree structure that accurately reflects the desktop application's user interface hierarchy. 

Within this tree structure, the nodes represent the control elements of the application, and the edges denote the actions performed on the parent node to access the child node. Take, for instance, a node that represents the "Margins" control element in Microsoft Word. The edge from its parent node, namely, the "Layout tab" control element,  is labeled by the action \textit{Click}, meaning that a click on the ``Layout tab'' control element reveals the``Margins'' control element, among others. The``Margins'' node has several outgoing edges to the children nodes, one child node per setting option. For example, under the``Margin''  node, ``Narrow'' is a child node and the label on its edge from the parent ``Margins''   is labeled 
\textit{Select Narrow}. 
 The path in the tree for this selection is: (Layout->Margins->Narrow). The user actions necessary to navigate and adjust the margin settings of the document is also captured on this path as metadata.

This combination of node and edge offers a comprehensive and hierarchical depiction of the tree of control elements and actions that can be executed on them (see ACT in Figure~\ref{fig:technical-diagram}).

\subsubsection{\textbf{Creating the Few-Shot Examples Dataset}}

\paragraph{\textbf{Pair generation.}}

The \textit{Pair Generator} module utilizes the previously built ACT to create  ~\controlvalue{} pairs for each of the application controls. For example, in \textit{Word}, the \textit{Editing} control element provides three choices: \textit{Editing}, \textit{Reviewing}, and \textit{Viewing}. These choices are the values for the \textit{Editing} control element, which will result in these three \controlvalue{} pairs to be created: <Editing, Editing>, <Editing, Reviewing>, <Editing, Viewing>.
However, some control elements might not have corresponding values; for instance, the \textit{Bold} control element in \textit{Word} does not provide any options. In such a case, the value for the \textit{ Bold} control element is set as \textit{ none}.

\paragraph{\textbf{Few-shot examples generation.}}

Recall from Section ~\ref{overview} on the overview that the \textit{Seed LLM} module generates the few shot examples dataset for the application and saves it in the FED. Each such example in the FED is a triple <NLC, CE, Value(CE)>. The semantics of the triple provide an interpretation for \command{}, namely, linking the action of assigning Value(CE) to the control element CE.

For this purpose, we provide prompts to the \textit{Seed LLM} to generate the triples in the FED (Few-Shot Example Dataset
- see Figure~\ref{fig:technical-diagram}). The prompt for \textit{Seed LLM} includes a few hand-crafted few-shot examples, some chosen from the WOZ study (see Section~\ref{woz}), and a ~\command{} corresponding to each example as a response (see \textit{Seed LLM} in Figure~\ref{fig:technical-diagram}). These hand-crafted prompts instruct the Seed LLM on the creation of the NLC for~\controlvalue{} pairs. Given a~\controlvalue{} pair originating from the \textit{Pair-Generator} module, 
the \textit{Seed LLM}  output is a ~\command{} corresponding to the <CE, Value(CE)> pair that gets stored in the FED. We utilized the \textit{GPT-4} model~\cite{achiam2023gpt} with few-shot learning, as the \textit{Seed LLM}, to produce \command{}s for each~\controlvalue{} pair for the application.

These few-shot examples were general and varied in several aspects. First, they covered a number of diverse applications, such as file management (\textit{Explorer}), text editing (\textit{Wordpad}, \textit{Notepad}, and \textit{Microsoft Word}), video conferencing (\textit{Zoom}), music streaming (\textit{Spotify}), and email (\textit{Outlook}). The diversity of applications ensured the model's ability to generalize across various computer applications and functionalities. Second, the examples included a broad range of command types, like creating new items (e.g., new folder), text formatting (e.g., strike-through), meeting management (e.g., initiating a new meeting with video off), text editing (e.g., cut), formatting adjustments (altering font size), playback controls (e.g., pause) and interface display (e.g., display the to-do bar). The breadth of command types ensured that the model was trained to generate a wide range of accurate \command{}s. The prompt snippet of few-shot examples along with the output from the \textit{Seed LLM} appears in the Appendix~\ref{subap:secondary-prompt}.
\textit{It is important to note that these hand-crafted prompts are generic and that the same set of few-shot examples is used to generate the FED dataset for any application}.

\paragraph{\textbf{Curating the FED dataset.}}

The FED consists of few-shot examples for $11$ different applications (see Appendix \ref{subap:applications}) and for all the controls of these applications. Smaller applications, such as \textit{Notepad}, provided roughly $50$ examples, whereas larger applications, like \textit{Excel}, offered close to $400$ examples. Each example was independently assessed by two members of our research team for the accuracy of the generated few-shot examples.  Examples found to be erroneous were discarded. For example, within the \textit{Excel} application, we came across a pair \controlvalue{} (coming soon, none), which the \textit{Seed LLM} was unable to translate into a pertinent natural language command and hence was discarded.
In a similar vein, in \textit{Word}, the \controlvalue{} pair <less, none> was linked erroneously to the command ``Reduce the size of the selected text,''
After cleaning the data set, we converged on $1,200$ high-quality few-shot examples in the FED.

\subsection{Run-Time Component}

The runtime component will interpret the user's \command{} by generating an accurate \controlvalue{} pair reflecting the interpretation, associate an Action on the \controlvalue{} pair, and execute the sequence of steps corresponding to the Action.

The \textit{Gen LLM} module is responsible for producing a \controlvalue{} pair that mirrors the interpretation of the user's \command{}. This process leverages the capabilities of GPT-4 with few-shot learning.  We describe this process next.

The prompt it uses comprises the user's NLC input along with a collection of few-shot examples that are pertinent to the NLC. The \textit{Action Relay} module generates the sequence of steps in the screen reader to execute the Action on the  \controlvalue{} pair. To this end, it references the information in ACT that facilitates execution of the step sequence.

\subsubsection{\textbf{Prompt Engineering}}

The prompt it uses comprises the user's NLC input along with a collection of few-shot examples that are relevant for user's input \command{}. Relevancy is calculated using the \textit{BERT} module.  Using BERT word embedding,  it computes the semantic similarity of user's \command{} and the few-shot example data set of the application in the FED. The result of the \textit{BERT} module is a set of relevant few shot examples for the user's \command{}.
For example, if the user command is ``Set the Margin to Narrow'', the semantic similarity search based on BERT embedding will pull all related examples linked with the \textit{Margin} control element such as ``Change the Margin to Narrow, (Margins, Narrow)'', ``Change Margin to Normal, (Margins, Normal)'', ``Change Margin to Moderate'', etc (see \textit{Gen LLM} part in Figure~\ref{fig:technical-diagram}). Using word embedding to retrieve relevant few-shot examples is crucial to ensure that \textit{Gen LLM} generates a \controlvalue{} pair that is an accurate reflection of the interpretation of user's \command{}.
  
% User's \command{} and the relevant few shot examples are used as prompts for \textit{Gen LLM} to generate a \controlvalue{} pair demoting the interpretation of the \command{}. A segment of the prompt that includes a few shot examples and the output for a sample user command appears in the appendix~\ref{subap:main-prompt}.

User's \command{} and the relevant few shot examples are used as prompts for \textit{Gen LLM} to generate a \controlvalue{} pair demoting the interpretation of the \command{}. A prompt containing several few shot examples and the output for a sample user command is included in the appendix~\ref{subap:main-prompt}.

\subsubsection{\textbf{Action Relay}}

The \textit{Action Relay} module generates the sequence of screen reader steps to execute the Action on the  \\
\controlvalue{} pair generated by \textit{Gen LLM}. To this end, it references the information in ACT that facilitates the execution of the step sequence.
 For example, when the NLC 'Change font size to 12' is processed, the \textit{Action Relay} module carries out the screen reader steps in the following order: \textit{Select Font Group Menu}, \textit{Select Font Dropdown}, and \textit{Select Font Size to 12}. Once the task is accomplished, the system provides a 'Font size updated 12' confirmation message to the user. If the action fails, the system prompts the user to re-issue the command and the process begins anew.

\subsubsection{\textbf{Handling Ambiguity}}

On occasions, a user's \command{} may be large.  For instance, the \textit{Notepad} application: If a user instructs ``erase the highlighted text'', they may be referring to either the \textit{delete} or \textit{cut} function. The {Gen LLM}  will need to present both of these options to the user for disambiguation.
Fortunately, the underlying \textit{GPT-4} driving \textit{Gen LLM} can be configured to provide multiple responses in rank order. A simple postprocessing step \sysname{} will present these choices to the user only when the responses differ.

\subsubsection{\textbf{Evaluation of Gen LLM}}

We used the earlier ground-truth WOZ study dataset to evaluate the performance of Gen LLM. For each example in the WOZ dataset, we fed the \textit{NLC} as input to our model to generate the output of the \controlvalue{} pair, which was then compared to the ground truth pair in the data set. The overall accuracy of our Gen LLM was $78\%$ ($113$ out of $145$ \textit{NLCs}).

We also analyzed the pattern where the model gave incorrect outputs. We noticed that incorrect outputs from \textit{Seed LLM} were primarily caused by natural language commands containing multiple instructions within a single command. For example, some of the tasks in our WOZ study were multi-step tasks, meaning they can be completed by giving a single \textit{composite} command or multiple single-action commands. For example, to move the dog picture to the photos folder, a participant can use the \textit{move} composite query or use \textit{Copy} and \textit{Paste} single action commands. However, many participants opted to include both \textit{Copy} and \textit{Paste} within a single command, which \textit{Gen LLM} was unable to process and generate the action. This limitation is discussed further in the Section~\ref{discuss}.

\section{User Study}

We conducted an IRB-approved user study with blind screen reader users to test the effectiveness of \sysname{} against the status quo. 

% Specifically, the study aimed to confirm the following hypotheses:
%  \begin{enumerate}
%      \item \textbf{H1:} Higher efficiency in accessing the desired controls.
%      \item \textbf{H2:} Less effort is required for locating and interacting with desired application content.
%      \item \textbf{H3:} The users are more efficient, i.e., complete tasks faster,
%      with~\sysname{} than with a regular screen reader.
%      \item \textbf{H4:} ~\sysname{} is more usable than a regular screen reader; desktop interaction is easier and more intuitive with~\sysname{} than with a regular screen reader.
%  \end{enumerate}

\renewcommand{\arraystretch}{1.5}
 \begin{table*}[t]
    \begin{adjustbox}{center}
    \begin{tabular}{|c|c|c|c|c|}
    \hline
         \textbf{ID} & \textbf{Gender} & \textbf{Age} & \textbf{Acuity} & \textbf{Screen Reader} \\ \hline
         \textbf{P1} & Male & $43$ & Light Perception & JAWS, NVDA \\ \hline
         \textbf{P2} & Male & $58$ & No Vision & JAWS, NVDA  \\ \hline
         \textbf{P3} & Male & $40$ & Light Perception & JAWS, NVDA, VoiceOver \\ \hline
         \textbf{P4} & Male & $36$ & No Vision & JAWS \\ \hline
         \textbf{P5} & Female & $39$ & No Vision & JAWS, NVDA, VoiceOver \\ \hline
         \textbf{P6} & Female & $64$ & Light Perception & JAWS \\ \hline
         \textbf{P7} & Female & $61$ & No Vision & JAWS \\ \hline
         \textbf{P8} & Female & $31$ & Light Perception & JAWS, VoiceOver \\ \hline
         \textbf{P9} & Female & $35$ & No Vision & JAWS, NVDA, VoiceOver \\ \hline
         \textbf{P10} & Male & $44$ & Light Perception & JAWS, NVDA, VoiceOver \\ \hline
         \textbf{P11} & Female & $35$ & Light Perception & JAWS, VoiceOver \\ \hline
    \end{tabular}
    \end{adjustbox}
    \setlength{\abovecaptionskip}{10pt}
    \captionsetup{justification=centering}
    \caption{User study participant demographics.}
    \label{tab:final_study_participants}
\end{table*}

\subsection{Participants} 
We recruited $11$ blind screen reader users to participate in the study. Recruitment was done by leveraging an email contact list followed by snowball sampling. The inclusion criteria were: (i) having a visual impairment severe enough to require screen readers for interacting with computers; (ii) proficiency with \textit{JAWS} screen reader software; and (iii) experience with the \textit{Windows} operating system and software applications. The average age of the participants was $44.2$ (median=$40$, min=$35$, max=$64$), and the gender representation was almost equal ($5$ male, $6$ female). All participants reported owning a desktop or laptop computer. Additionally, they were all very comfortable using the \textit{JAWS} screen reader software, with $6$ participants also having experience in using \textit{NVDA} and $6$ using \textit{VoiceOver}. More self-reported details of the participant demographics can be found in Table~\ref{tab:final_study_participants}. There was no overlap between the participant pools of the earlier WOZ study and this study, thereby promoting external validity.

\subsection{Study Design} 

In the study, participants had to complete typical computer tasks in $6$ different applications: \textit{Amazon} (opened in \textit{Firefox}), \textit{Excel}, \textit{File Explorer}, \textit{Google Docs} (opened in \textit{Chrome}), \textit{Outlook}, and \textit{Spotify}. Each of these applications represented a unique use case (e-commerce, productivity, text editing, e-mailing, etc.) to evaluate the generalizability of our system. Exact details of the tasks assigned for each application are shown in Table \ref{tab:final_study_subtask}. We selected these tasks as representative and useful for screen reader users.

The study had a within-subject design where participants interacted with an application under $3$ conditions. These conditions were: 
\begin{itemize}
    \item \textbf{Screen Reader (JAWS) only:} Participants could only use JAWS screen reader software.
    \item \textbf{Screen Reader with a state-of-the-art \syscomp{} interface:} Participants were allowed to use a state-of-the-art \syscomp{} system~\cite{uckun2022taming} that (as explained in Related Work) enabled screen reader users to access all the \textit{visible} controls via a custom pop-up interface. 
    \item \textbf{Screen Reader with ~\sysname{}'s command interface:} Participants were allowed to use \sysname{}'s natural language interface to do the assigned tasks.
\end{itemize}

\textbf{Note:} AccessBolt and Savant differ in their interface and accessibility approaches. AccessBolt utilizes a pop-up interface that displays all visible control elements, requiring users to manually navigate and select these controls to perform tasks. Consequently, interaction is confined to visible control elements only, requiring users to manually reveal invisible controls for AccessBolt to access them. In contrast, Savant offers a natural language interface, allowing users to execute tasks through natural language commands (NLC), providing a more intuitive and seamless interaction experience that doesn't restrict users to visible controls only.

Participants interacted with $2$ different applications under each condition. The ordering of the conditions and applications across participants was counterbalanced using the Latin square method~\footnote{\url{http://compneurosci.com/wiki/images/9/98/Latin_square_Method.pdf}}.

\subsection{Apparatus}
The study was conducted in person, where the participants were provided with a Lenovo ThinkPad Windows laptop, which had all the necessary applications installed, including the \textit{JAWS} screen reader. An externally accessible QWERTY desktop keyboard was also connected to the laptop, as all participants mentioned that they were familiar with the standard keyboard during the recruitment process.

\subsection{Procedure}
The experimenter first explained the purpose of the study to the participants and obtained their informed consent. Then, the participants were given $20$ minutes to practice/revise their \textit{JAWS} skills, configure JAWS to best match their preferences, and get comfortable using the \syscomp{} and \sysname{} assistive technologies. For the practice session, we used the \textit{WordPad} application as the context, where the experimenter showcased the range and flexibility of command usage in \sysname{} to the participant. After the practice session, the participants we asked to complete the tasks under different conditions in the predetermined counterbalanced order. If the participant could not finish a task within $10$ minutes, it was considered a failure. After each task, an SEQ \footnote{\url{https://measuringu.com/seq10/}} question was asked, and after completing all the tasks, the NASA-TLX questionnaire~\cite{hart1988development} was administered. Lastly, the experimenter concluded the study with an open-ended exit interview where the participant was asked to provide subjective feedback, including feature requests and suggestions for improvement. With the participant's permission, recording features were active throughout the study to capture all interaction activities for post-study analyses. Each study lasted for nearly $2$ hours. All conversations were in English, and the participants were compensated monetarily for their time and contribution.

\subsection{Measurements}

From the study, we collected the following metrics: (i) the average time spent on each task; (ii) the average number of shortcuts pressed per task, including navigation; (iii) the scores from the Single-Ease-Questionnaire (SEQ) on a 7-point Likert Scale, where 1 represents 'Hard' and 7 'Easy'; and (iv) NASA-TLX questionnaire responses for each condition. The SEQ was used to assess the perceived difficulty of each application, while the NASA-TLX questionnaire was administered after each condition to gauge the participants' perceived workload. We used qualitative analysis methods for the textual data collected by transcribing the participants’ utterances and qualitative feedback from the exit interviews. Specifically, we used the standard open coding and axial coding techniques~\cite{saldana2021coding} to iteratively annotate the textual data and excavate key insights and themes recurring in the data. We detail our findings next.

% \fixmevk{Refer to past papers and mention qualitative exit-interview data and how you analyzed this data using open coding.}

% (iii) the frequency of back-and-forth navigation during a task;~\fixmevk{not well defined, and also not included in results} 

% \renewcommand{\arraystretch}{1.5}
% \begin{table*}[t]
%     \begin{adjustbox}{center}
%     \begin{tabular}{|c|c|c|} \hline
%         \textbf{Application} & \textbf{Task 1} & \textbf{Task 2} \\ \hline
%         Amazon & Go to the Best Sellers tab (depth=2) & Change the language of the website (depth=2) \\ \hline
%         Excel & Insert a pie chart (depth = 3) & Add a new note (depth = 2)\\ \hline
%         File Explorer & Open the PDF document & Move the dog.png \\ 
%         & with Adobe Acrobat (depth = 2) & to the Pictures folder (depth = 3)\\ \hline
%         Google Docs & Change font size to 14 (depth = 1) & Insert an image (depth = 3) \\ \hline
%         Outlook & Flag this email (depth = 1) & Arrange emails by date (depth = 2)\\ \hline
%         Spotify & Create a new playlist (depth = 2) &  Start a private listening session (depth = 2) \\ \hline
%     \end{tabular}
%     \end{adjustbox}
%     \setlength{\abovecaptionskip}{10pt}
%     \captionsetup{justification=centering}
%     \caption{Tasks the participants had to complete in the user study. Depth represents the number of user interactions needed to navigate from the root to a specific element to complete the task.}
%     \label{tab:final_study_subtask}
% \end{table*}

\renewcommand{\arraystretch}{1.5}
\begin{table*}[t]
    \begin{adjustbox}{center}
    \begin{tabular}{|c|c|c|c|c|} \hline
        \textbf{Application} & \textbf{Task 1 Description} & \textbf{Task 1 Depth} & \textbf{Task 2 Description} & \textbf{Task 2 Depth} \\ \hline
        Amazon & Go to the Best Sellers tab & $2$ & Change the language of the website & $2$ \\ \hline
        Excel & Insert a pie chart & $3$ & Add a new note & $2$ \\ \hline
        File Explorer & Open the PDF document & \multirow{2}{*}{$2$} & Move the dog.png & \multirow{2}{*}{$3$} \\
        & with Adobe Acrobat & & to the Pictures folder & \\ \hline
        Google Docs & Change font size to 14 & $1$ & Insert an image & $3$ \\ \hline
        Outlook & Flag this email & $1$ & Arrange emails by date & $2$ \\ \hline
        Spotify & Create a new playlist & $2$ & Start a private listening session & $2$ \\ \hline
    \end{tabular}
    \end{adjustbox}
    \setlength{\abovecaptionskip}{10pt}
    \captionsetup{justification=centering}
    \caption{Tasks the participants had to complete in the user study. Depth represents the number of user interactions needed to navigate from the root to a specific element to complete the task.}
    \label{tab:final_study_subtask}
\end{table*}

\subsection{Results}

\subsubsection{\textbf{Completion Rate}}

% Out of the $18$ failed tasks under the JAWS condition, $13$ was due to participants giving up on the task midway, citing frustration as the cause. % Out of $22$ tasks that participants interacted with in the JAWS condition, only $4$ were successfully completed ($\sim 18\%$ success rate). 

% \fixmevk{give example from the study task}. 

The completion rate varied across study conditions. In the JAWS condition, the participants completed only $6$ out of the $22$ tasks. The completion rate in the~\syscomp{} condition was much higher ($14$ out of the $22$ tasks), however it was the highest in the \sysname{} condition ($18$ out of $22$ tasks). A closer analysis of the user behavior during the tasks revealed that in the JAWS condition, one of the main reasons the participants struggled to complete the tasks was because of \textit{invisible} controls that could be accessed only by opening up the corresponding `container' control. For instance, participants had difficulty locating the \textit{Date} sorting option in \textit{Outlook}. This option was nested within the \textit{Sort By} sub-menu, found under the \textit{View Menu}. Specifically, in $8$ of the $16$ completion failures in the JAWS condition, the participants tried to locate the control by sequentially navigating over the visible GUI elements on the screen without realizing that they had to first select a parent control (e.g., View Menu) before they could access the desired target control (e.g., Sort By). In these $8$ cases, only after trying for some time and failing to locate the target control the participants explore other strategies, such as exploring menus, sub-menus, and tabs, but they ran out of time before they could find the target control. A similar pattern was observed in the \syscomp{} condition, where in $6$ out of $8$ completion failures, the participants unsuccessfully tried to locate the target control in its pop-up interface, and eventually ran out time when they were trying to find the control in the application's GUI. The few failures in the \sysname{} condition were all due to the LLM errors where user commands could not be accurately mapped to the desired target control, despite multiple attempts by the participants.    

% In the study, a significant number of participants struggled to complete the tasks, primarily because many of them required participants to interact with multiple invisible controls. Specifically, out of the $18$ failed tasks under the JAWS condition, $13$ participants made concerted efforts to locate the correct invisible control by sequentially navigating through the control elements. Ultimately, they abandoned the task due to frustration. These participants spent $258$ seconds ($\sim 4$ minutes) on average before stating they could not complete the tasks. In the \syscomp{} condition, participants were required to locate the controls within an alternative pop-up interface. Because the system did not directly display the invisible controls, participants had to select a visible control element to reveal the invisible controls. This multi-step process was time-consuming and challenging, leading many participants to struggle to find the correct control within the allotted time. In the \sysname{} condition, participants couldn't complete the task primarily because the LLM didn't produce the correct response, likely due to the complexity of the provided natural language command.

\subsubsection{\textbf{Average time spent per task}}

\begin{figure*}   
    \centering
    \captionsetup{justification=centering,margin=2cm}
    \begin{subfigure}[t]{0.49\textwidth}
        \includegraphics[width=\linewidth]{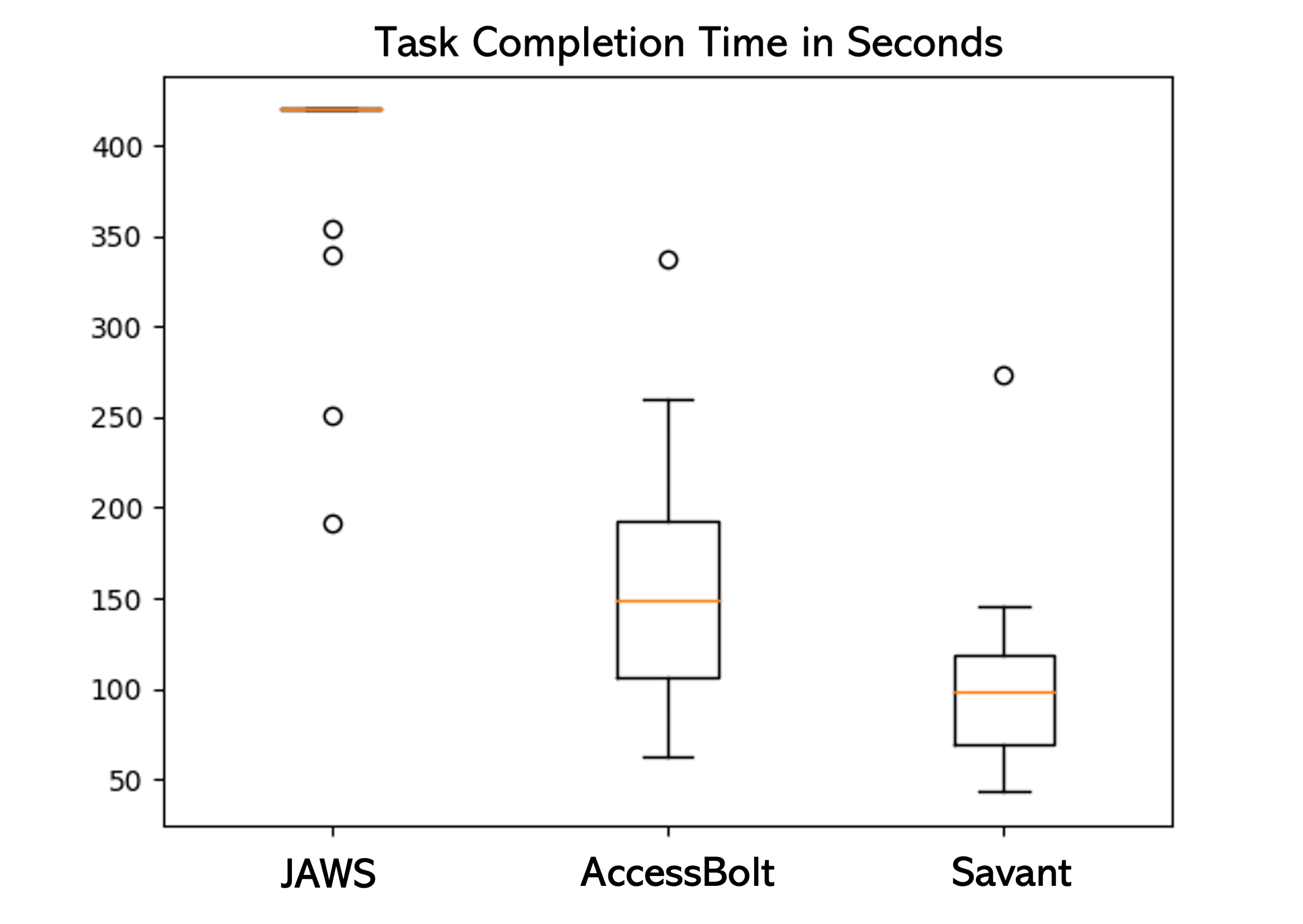}
        \caption{Box plot showing task completion times in seconds.}
        \label{fig:completion_time_boxplot}
    \end{subfigure}
    \begin{subfigure}[t]{0.49\textwidth}
        \includegraphics[width=\linewidth]{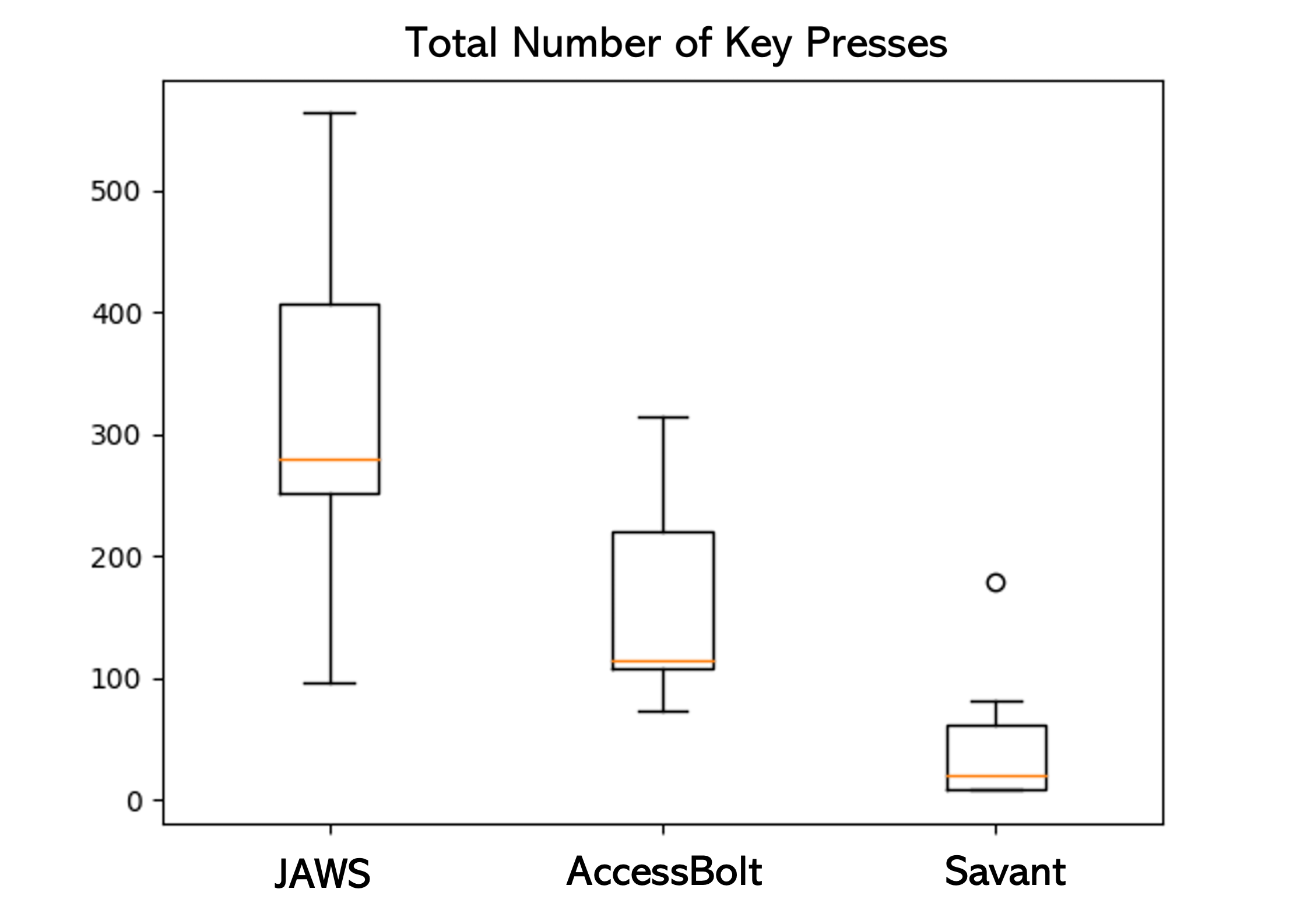}
        \caption{Box plot showing the total key presses recorded under each condition.}
        \label{fig:key_press_boxplot}
    \end{subfigure}
    \caption{Box plots for quantitative measurements collected during the user study.}
\end{figure*}

On average, the participants spent $395$ seconds under JAWS condition (median=$420$, SD=$61.1$), $159$ seconds under~\syscomp{} condition (median=$148.5$, SD=$69.6$), and $101$ seconds under~\sysname{} condition (median=$98$, SD=$49.5$) to do the tasks. Figure~\ref{fig:completion_time_boxplot} shows the box plot for the task completion times for the three conditions. A Shapiro-Wilk test showed that the data for JAWS (W=$0.478$, p=$8.15e^{-8}$) and~\sysname{} deviated from normal distribution (W=$0.821$, p=$0.001$) whereas~\syscomp{} data had normal distribution (W=$0.944$, p=$0.236$). First, we conducted a Wilcoxon Signed Rank test between JAWS and~\syscomp{}, which showed that the difference between the completion times was statistically significant (W=$0.0$, p=$4.0e^{-5}$). Second, we ran the same test between~\syscomp{} and~\sysname{} conditions, and the result showed statistical significance between these two conditions as well (W=$51.5$, p=$0.01$).

As explained previously, in the JAWS and~\syscomp{} conditions, the participants spent considerable time searching for the target control in the application's GUI, which was especially the case for `invisible' controls. The longest duration a participant persisted was $7$ minutes before giving up. In contrast, the~\sysname{} condition eliminated this search process; participants could simply issue a natural language command, and access to the control was automated. Most of the time spent in the~\sysname{} condition was allocated to processing the natural language command by the LLM, with a portion dedicated to executing the entire pipeline. In some cases, the participants also had to rephrase or repeat the command because the LLM was unable to correctly attribute the command to the corresponding application control (due to speech-recognition errors as well as ambiguity in utterances).
Occasionally, due to ambiguity in the given natural language command, the system presented multiple options to the user. In these instances, participants spent additional time determining the correct control element.

% ~\fixmevk{I don't think we have this dialog feature in the architecture section}.

\subsubsection{\textbf{Key Press Statistics}}

On average, the participants pressed $319$ keys under JAWS condition (median=$279$, SD=$132.8$), $163$ keys under~\syscomp{} condition (median=$115$, SD=$189.1$) and $44$ under~\sysname{} condition (median=$20$, SD=$51.9$). Shapiro-Wilk test showed divergence from a normal distribution for~\syscomp{} (W=$0.83$, p=$0.25$) and~\sysname{} conditions (W=$0.75$, p=$0.001$). Figure \ref{fig:key_press_boxplot} shows the boxplot for the total number of keys pressed. A Wilcoxon Signed Rank test between JAWS and~\syscomp{} conditions showed the reduction in key presses is statistically significant (W=$2.0$, p=$0.006$). Another Wilcoxon test between JAWS and~\sysname{} conditions also showed that reducing key presses was statistically significant (W=$0.0$, p=$0.003$).

% \fixmevk{What about between AccessBolt and Savant? Also, for completion failure, what value of shortcuts was given, can we say the value at the end of 10 mins?}

% Furthermore, key press data were analyzed to detect the number of navigation keys (Tab and arrow keys) participants pressed rather than typing key presses. These analyses showed that under the JAWS condition, $91\%$ of total key presses were navigation keys while this percentage was $70\%$ for~\syscomp{} and $30\%$ for ~\sysname{} conditions. 

Furthermore, key press data were analyzed to determine the proportions of basic navigation keys (Tab and Arrow keys) used by the participants within the total key presses during the tasks. The analysis revealed that in the JAWS condition, a striking $91\%$ of the total key presses were these basic navigation keys. In contrast, this percentage decreased to $70\%$ in the~\syscomp{} condition and dropped even further to $30\%$ in the~\sysname{} condition. This shift in key press distribution underscores the differences in interaction behavior across the three conditions. In the JAWS and~\syscomp{} conditions, participants predominantly relied on navigation keys to search for the target control either within the application GUI or the \syscomp{}'s popup interface. On the other hand, in the \sysname{} condition, the use of navigation keys was naturally lesser due to the availability of natural language input alternatives.

% \fixmevk{I like this detail, but it is little weird that we dont explain what the rest 70\% was in the \sysname{} condition}

\subsubsection{\textbf{NLC Analysis}}

% On average, participants issued $4$ commands per task, with a total of $88$ natural language commands (NLCs) collected during the study. 

Out of $22$ tasks assigned to the participants in the \sysname{} condition, $11$ were successfully completed on the first attempt (i.e., the LLM correctly identified the~\controlvalue{} pair), while $7$ were completed after multiple tries. This repeated interaction contributed to longer completion times. $4$ tasks remained incomplete despite repeated attempts by the participants. Upon analyzing the cases where the system produced incorrect responses, two reasons were identified. Firstly, some user commands were complex (e.g., ``Select this cell and add a note''), consisting of multiple instructions within a single command. These commands often required several steps and involved accessing multiple controls. Our system is presently not equipped to handle a complex command and access multiple controls in one shot, leading to errors in execution. Secondly, errors in automatic speech-to-text conversion occasionally occurred, preventing the system from accurately interpreting the command and necessitating the user to repeat the command. 

\subsubsection{\textbf{Single Ease Question Score}}

We used the standard Single-Ease Question (SEQ) score to assess usability and to determine the perceived difficulty of tasks. The SEQ questionnaire asks participants to rate on a scale of $1$ to $7$, with $1$ being hard and $7$ being easy. On average, the tasks got a rating of $2$ under JAWS condition (median=$2$, SD=$1.19$), $5.6$ under~\syscomp{} (median=$6$, SD=$1.61$) and $6.57$ under~\sysname{} (median=$7$, SD=$0.80$). Shapiro-Wilk tests showed a departure from normality for all the conditions. A Wilcoxon test between JAWS and~\syscomp{} showed that the improvements in user experience were statistically significant (W=$2.0$, p=$7.36e^{-5}$). Another Wilcoxon test between~\syscomp{} and~\sysname{} also showed statistical significance (W=$15.0$, p=$0.01$).

\subsubsection{\textbf{NASA-TLX Score}}

We administered the standard NASA-TLX questionnaire~\cite{hart1988development} to gauge the workload experienced by the participants while they did the assigned tasks in different study conditions. NASA-TLX generates a score between 0 and 100 to estimate perceived task workload, where lower TLX values indicate lesser workloads and, hence, better performance. Overall, we observed that the study conditions had a significant impact on NASA-TLX scores. The TLX score for~\sysname{}(mean=$16.16$, median=$15.0$, SD=$8.57$)) condition was significantly lower than for~\syscomp{} (mean=$28.93$, median=$29.16$, SD=$5.09$) and JAWS (mean=$49.9$, median=$45.8$, SD=$20.9$) conditions. A Wilcoxon signed-rank test between JAWS and~\syscomp{} conditions showed that the difference in TLX scores was statistically significant (W=$2$, p=$.005$). Another Wilcoxon test between~\syscomp{} and~\sysname{} also showed statistical significance (W=$2.5$, p=$.006$). Friedman's test for NASA-TLX produced a chi-square statistic of $18.18$ (df = $2$) and a p-value of $0.00011$, implying statistical significance at p < $.05$.

A closer inspection of the responses to individual subscales of the TLX questionnaire revealed that the Effort and Frustration subscales contributed more toward the relatively higher workload scores in the JAWS condition, i.e., the ratings for these subscales were significantly higher than those for the other subscales (Temporal Demand, Physical Demand, Mental Demand, and Overall Performance). For the~\syscomp{} condition, Effort was the discriminating subscale that received much higher ratings than the other subscales. The ratings for~\sysname{}, however, were uniform across all the six subscales.

\subsubsection{\textbf{Qualitative Feedback}}

A few notable observations that were uncovered from the exit interviews are listed next.

\textbf{Participant preferences and feedback on~\sysname{}.} All participants expressed their preference for~\sysname{} when asked about their preferred system. Participants $P1$, $P2$, $P3$, $P6$, and $P11$ praised Savant for its high user-friendliness. Participant $P2$, for instance, said, ``I like the voice (~\sysname{}) because sometimes I feel lazy, and using the voice is direct and easy.'' Meanwhile, Participants $P4$, $P7$, $P8$, and $P9$ highlighted that they appreciated having fewer keyboard presses and simpler navigation to find the controls. According to $P8$, ``Less navigation, fewer keyboard presses make everything faster.'' Participants $P6$ and $P9$ also pointed out that they liked how ~\sysname{} had the same command interface for all applications, which they thought was a cool idea. $P6$ said, ``People (screen reader users) are used to Tab and Shift+Tab controls. Performing a task with only a few familiar keys is much better.'' P5 stated that our system should have an alternate pop-up interface like~\syscomp{} Their reason was: ``It (keyboard interface) gives me more opportunities when I can't use my voice: in a silent place (library), in a Zoom meeting, when there are background noises''. 

% I don't have to memorize the shortcuts''. P5 stated that they liked the keyboard interface of the ~\sysname{}. Their reason was: ``It (keyboard interface) gives me more opportunities when I can't use my voice: in a silent place (library), in a Zoom meeting, when there are background noises''. 

\textbf{Suggestions for improvement.} Two participants, $P1$ and $P3$, suggested that \sysname{} could be improved by supporting more complex commands such as composite commands. To quote $P1$, ``It (~\sysname{}) should support commands like open the document and copy the text''. $P5$ and $P9$ suggested allowing users to change the keyboard shortcut that triggers the system or activate the command interface. Lastly, $P1$ and $P4$ suggested that our system should have a wake word for triggering instead of a push-to-talk button.

% When asked how~\sysname{} could be improved, participants $P1$ and $P3$ suggested support of more complex commands such as composite commands. To quote $P1$, ''It (~\sysname{}) should support commands like open the document and copy the text''. $P5$ and $P9$ suggested allowing users to change the keyboard shortcut that triggers the system or activate the command interface. Lastly, $P1$ and $P4$ suggested that our system should have a wake word for triggering instead of a push-to-talk button.

\textbf{Tedious to work with keyboard shortcuts.} When asked whether they had any problems with keyboard shortcuts used in computer applications, participants $P3$, $P8$, $P9$, and $P10$ complained about the heterogeneity of keyboard shortcuts used across computer applications and operating systems; $P10$ said, ``Navigating a web page uses certain commands. When in an application, there are other generic controls and application-specific controls. There should be a universal guide for app developers'', while $P9$ mentioned, ``Sometimes shortcuts change depending on what you are doing. You need to be a power user to know them all''. Some of the frustrations participants sounded were: $P4$: ``I can't memorize (shortcuts), I have to go look it up in forums or ask email groups'', $P5$:``You are busy with content. Then, you have to format the content, and it gets you stressed. You lose focus. Get distracted from the task''. One point all participants agreed upon was that they had to keep practicing keyboard shortcuts or that they had forgotten how to do specific tasks.
\section{Discussion}

\label{discuss}

The results of our user study are indicative of the potential of \sysname{} to tackle the interaction challenges faced by blind computer users, arising from the diversity of the interfaces of applications. 

\subsection{Limitations and Future Work}

Here, we discuss the limitations of \sysname{} and a roadmap for handling them.

\subsubsection{\textbf{Supporting Pop-up and Sub-windows Controls}}

In the qualitative feedback, numerous participants pointed out the advantages of enabling \sysname{} to manage tasks involving pop-ups and sub-windows. To enhance our system with these features, several challenges must be tackled. First, such windows manifest unpredictably, complicating the prediction of their emergence and the controls they include. Second, a comprehensive analysis of the relationships between various pop-ups/sub-windows and the primary window is essential for extracting all application controls. Third, the complexity of the extraction process increases when multiple actions on control elements trigger the appearance of identical pop-up windows. Additionally, the controls within these pop-ups and sub-windows often show context-dependent behaviors, which poses significant difficulties in their programmatic extraction and interaction. For instance, the options on the toolbar can vary depending on whether the user is editing text, adding a table, or adjusting image formats. Addressing these issues will be the focus of future development.

\subsubsection{\textbf{Supporting Complex Commands}}

Currently, ~\sysname{} is capable of executing simple single-action commands such as button clicks or selections from a drop-down menu. Feedback from participants in our study suggests that ~\sysname{} should be enhanced to handle more complex commands. For instance, the command 'Make all text italic' would need to be separated into two simpler actions: \textit{Click Select All button} and \textit{Click italic button}. To address such complex commands, a task decomposition model similar to the one in \cite{zhang2021hierarchical}, could break down complex commands into single-action tasks for ~\sysname{} to execute.

\subsubsection{\textbf{Supporting Complex Task Involving Multiple Applications}}

Many computer tasks in practical situations involve multiple control elements (e.g., \textit{load the (named) file from the Documents folder into the editor}), possibly across multiple applications (e.g., \textit{Microsoft Word} and \textit{Explorer}). Automating such tasks will require reasoning over potentially multiple applications' contexts, identifying application-specific intent underlying the request within each application, and seamlessly stitching together numerous user-interface operations in appropriate execution sequences. It would be possible to use LLMs for this task, a topic for future research.
% We can potentially leverage the power of LLMs with a few-shot learning method to guide the LLMs to output both the application names and intents given an input request and the applications' contexts.  The intents, application names, and applications' non-visual GUI representations packaged as either a tree of thoughts~\cite{yao2023tree} or a graph-of-thoughts~\cite{besta2023graph} and a few guiding examples will all be condensed into a prompt, and the user request will be posed as a query.

\subsubsection{\textbf{Applicability to complex GUIs and sequential tasks}}

Extending our LLM-based method to complex sequential tasks is possible using a Chain-of-Thought (COT) prompting strategy~\cite{wei2022chain}. This strategy helps LLMs handle complex tasks by breaking them into logical steps. For a user's natural language command~\textit{C}, the prompt~\textit{P} will include chain-of-thought in-context few-shot examples guiding the LLM in generating an accurate action sequence.

% For example, the prompt to create a ``table with $2$ rows and $3$ columns'' will include the NLC, subtasks for inserting tables and selecting rows and columns, and actions: [(Insert, Select), (Table, Select), (Insert Table, Select), (No of Rows, $2$), (No of Columns, $3$), (OK, Select)]. This guides the LLM in breaking down tasks into smaller steps. Thus, for the query ``set text alignment to justified and line and paragraph spacing to $1.5$'', the prompt will decompose it into text alignment and spacing adjustment, generating the actions: [(Text Alignment, Justify), (Line Spacing and Paragraph, 1.5)].

\subsubsection{\textbf{Multilingual support and future extensions}}

Large language models (LLMs) demonstrate remarkable capabilities in understanding and translating instructions from non-English languages into English. This proficiency suggests a promising avenue for enhancing~\sysname{} by integrating translation functionalities, enabling support for multiple languages. Such an expansion could significantly broaden the accessibility and usability of~\sysname{} for non-English speaking users, presenting an exciting direction for future research.
% Adding translator\footnote{\url{https://pypi.org/project/googletrans/}} functionality could not only streamline the interaction process for a diverse user base but also align with global accessibility standards, making \sysname{} a more inclusive technology.}

\subsubsection{\textbf{Wake-word Implementation}}

The results of the user study indicated a preference for activating the system with a \textit{wake-word} instead of using a push-to-talk button. Participants noted that a system not dependent on keyboard input could greatly assist people unable to use keyboards due to various disabilities (such as amputation, paralysis, or motor impairments). Implementing this feature requires a continuously active microphone and a wake-word. Deploying wake-word detection in desktop applications involves several challenges, such as the need for accurate detection, instant processing, addressing privacy issues, and ensuring multi-platform compatibility. Overcoming these challenges requires thorough model training, optimization techniques, and effective management of privacy concerns.

% \subsection{Societal Impact}

% The importance of desktop usability cannot be overstated when it comes to guaranteeing equal access to digital material for people with disabilities, including those with significant visual impairments. Desktop application interfaces (GUI) are frequently developed with a primary focus on users with vision, which might result in usability difficulties for individuals who depend on screen readers or assistive devices. By harnessing the capabilities of Large Language Models (LLMs) like GPT-4, this approach promotes digital inclusivity. It empowers blind users to navigate and interact with various software applications more seamlessly, breaking down barriers to information, education, and employment. This advancement enhances the independence of blind users, reducing their reliance on complex and often disparate assistive technologies. It fosters a sense of autonomy, allowing them to participate more fully in the digital age. Moreover, it encourages developers and technology companies to prioritize accessibility, paving the way for a more inclusive digital landscape.
\section{Conclusion}

In this paper, we addressed the usability issues faced by blind screen reader users with desktop applications. From prior studies, it was evident that blind users face several issues while interacting with computer applications, such as difficulty in memorizing keyboard shortcuts, accessing invisible controls, etc. We proposed~\sysname{} interface that can automatically access and invoke all the application controls on the user's behalf by leveraging large language models (LLM), thereby allowing users to perform tasks uniformly across arbitrary desktop applications. By employing a design process that incorporated user feedback, we have successfully showcased notable improvements in the efficiency of computer interactions and the usability of desktop applications. This progress signifies a favorable advancement towards creating a digital environment that is more inclusive for individuals with visual impairments.
\section*{Acknowledgments}

We thank the anonymous reviewers for their insightful feedback. This work was supported by the Google Inclusion Research Award, NSF Awards:  2153056, 2125147, 2113485, and the following NIH Awards: 
R01EY030085, R01HD097188, and R01EY035688.

\bibliographystyle{ACM-Reference-Format}
\bibliography{bibliography}

\clearpage

\appendix
\section{Appendix}

\subsection{Example Prompt for the Seed LLM}\label{subap:secondary-prompt}

\begin{center}
    
\noindent\fbox{%
    \parbox{0.46\textwidth}{%

\textit{Comment: Diverse Applications Few-shot guiding examples}

\textbf{App:} File Explorer

\textbf{Control-Value pair:} (new item, folder)

\textbf{Response:} Create a new folder.\\

\textbf{App:} Wordpad

\textbf{Control-Value pair:} (strikethrough, none)

\textbf{Response:} Strike the selected text. \\

\textbf{App:} Zoom

\textbf{Control-Value pair:} (start a new meeting with video off, none)

\textbf{Response:} Start a new meeting without the video.\\

\textbf{App:} Notepad

\textbf{Control-Value pair:} (edit, cut)

\textbf{Response:} cut the selected text\\

\textit{Comment: Generating more examples}

\textbf{App:} Word

\textbf{Control-Value pair:} (share, copy link) \\

\textit{Comment: LLM-generated output}

\textbf{Response:} Share the document as a link.

    }%
}
\end{center}

% Prompt design for the Gen LLM
\subsection{Example Prompt for the Gen LLM}\label{subap:main-prompt}

\begin{center}
    
\noindent\fbox{%
    \parbox{0.46\textwidth}{%

\textit{Comment: Relevant Few-shot guiding example}

\textbf{Command:} Change the Margin to Narrow.

\textbf{Response:} (Margins, Narrow) \\

\textbf{Command:} Change the Margin to Normal.

\textbf{Response:} (Margins, Normal) \\

\textbf{Command:} Change the Margin to Moderate.

\textbf{Response:} (Margins, Moderate) \\

\textbf{Command:} Change the Margin to Wide.

\textbf{Response:} (Margins, Wide) \\

\textbf{Command:} Change the Margin to Mirrored.

\textbf{Response:} (Margins, Mirrored) \\

\textit{Comment: User request}

\textbf{User:} Set the Margin to Narrow. \\

\textit{Comment: LLM-generated output}

\textbf{Response:} (Margins, Narrow)
    }%
}
\end{center}

\subsection{Wizard-of-Oz-study}\label{subap:wizard-of-oz-study}

\begin{table*}[t]
    \begin{adjustbox}{center}
    \captionsetup{justification=centering}
    \centering
     \begin{tabular}{|c|c|c|c|c|c|}
        \hline
         \textbf{ID} & \textbf{Gender} & \textbf{Age} & \textbf{Acuity} &\textbf{Voice Assistant}& \textbf{Use Cases}\\ \hline
         \textbf{P1} & Male & $40$ & No vision & Alexa and Siri & Calendar, Email, Messages, and Reminder \\ \hline
         \textbf{P2} & Male & $58$ & No vision & GA & Calls, Email, and Messages \\ \hline 
         \textbf{P3} & Female & $64$ & Light perception & Alexa and Siri & Info Query and Timers \\ \hline
         \textbf{P4} & Female & $61$ & No vision & Siri & Alarms, Calls, Email, Info Query, and Messaging \\ \hline
         \textbf{P5} & Female & $39$ & No vision & Alexa, GA, and Siri & Calculation, Games, and Info Query \\ \hline
         \textbf{P6} & Female & $35$ & Light perception & Alexa and Siri & Alarms, Calling, Emails,  Start Apps, and Timers \\ \hline
         \textbf{P7} & Male & $43$ & Light perception & Alexa, GA, and Siri & App Navigation, Info Query, and Online Shopping \\ \hline
         \textbf{P8} & Male & $71$ & No vision & Alexa and Siri & Calls, Calculation, and Contacts \\ \hline
         \textbf{P9} & Female & $73$ & No vision & Alexa & Calculations and Info Query \\ \hline
         \textbf{P10} & Male & $62$ & Light perception & Alexa and GA & Alarms, Online Shopping, Smart Home, and Timers \\ \hline
         \textbf{P11} & Female & $31$ & Light perception & Alexa and Siri & Calls, Info Query, Online Shopping, and Timers \\ \hline
    \end{tabular}
    \end{adjustbox}
    \setlength{\abovecaptionskip}{10pt}
    \caption{Wizard-of-Oz study participant demographics. GA stands for Google Assistant. All information was self-reported.}
    \label{tab:voice-study-participant}
\end{table*}

\begin{table*}[t]
    \begin{adjustbox}{center}
    \captionsetup{justification=centering}
    \centering
    \begin{tabular}{|c|c|}
    \hline
         \textbf{Application}& \textbf{Task} \\ \hline
         Excel & Add a sum function to this cell \\ \hline
         Excel & Freeze the top row \\ \hline
         Excel & Change the format of this number to fraction \\ \hline
         Excel & Highlight the selected cell yellow \\ \hline
         File Explorer & Move dog picture to photos folder \\ \hline
         File Explorer & Create a new folder named ``Tasks''\\ \hline
         File Explorer & Sort files in folder \\ \hline
         File Explorer & Select all files \\ \hline
         File Explorer & Share selected files \\ \hline
         File Explorer & Open file with Adobe Acrobat \\ \hline
         Gmail & Find the email from Google \\ \hline
         Gmail & Move email to promotions folder \\ \hline
         Gmail & Go to starred emails folder \\ \hline
         Gmail & Reply to email saying ``Thank you'' \\ \hline
         Gmail & Refresh your inbox \\ \hline
         Word & Grow font size to $14$ \\ \hline
         Word & Make text bold \\ \hline
         Word & Center the text \\ \hline
         Word & Undo action \\ \hline
         Word & Switch style to heading $1$ \\ \hline
         Word & Add a comment saying "change later" \\ \hline
         Word & Insert a table with $2$ rows and $3 $columns \\ \hline
         Zoom & start a new meeting with video off \\ \hline
         Zoom & Copy the invitation link to the meeting \\ \hline
         Zoom & Schedule a meeting for tomorrow $1$ PM \\ \hline
         Zoom & Message your contact Alex saying ``Hello'' \\ \hline
    \end{tabular}
    \end{adjustbox}
    \setlength{\abovecaptionskip}{10pt}
    \caption{Tasks given to the participants in the Wizard-of-Oz study.}
    \label{tab:tasks}
\end{table*}

\subsubsection{\textbf{Participants}} We recruited $11$ ($P1$ to $P11$) screen reader users for our study. The average age was $52.5$ (median=$58$, SD=$15.1$, range=$31$-$73$). Gender representation was almost equal, with $6$ Female and $5$ male participants. Recruitment was done using an email list with simple random sampling. Inclusion criteria for the study were: (i) exclusive dependence on a screen reader with no functional vision, (ii) proficiency with \textit{JAWS} screen reader software, and (iii) familiarity with \textit{Windows} operating system platform and desktop applications. 

Participants were asked about their voice assistant usage habits during a pre-study interview. Almost all of the participants ($10$) mentioned that they used voice assistant systems at least once a day. $P2$ said he uses them at least once a week. $8$ participants mentioned that they use use Siri~\cite{Siri}, $9$ participants mentioned Amazon Alexa~\cite{Alexa}, and $4$ participants said they use Google Assistant~\cite{Google-Assistant}. Some common use cases for voice assistants were setting up alarms/timers/reminders, dictating messages and emails, making phone calls, or querying information such as weather, traffic, or trivia. More details about participant demographics and their voice assistant usage habits can be found in Table~\ref{tab:voice-study-participant}.

\subsubsection{\textbf{Design}}
During the study, participants were given tasks from $5$ applications: \textit{Excel}, \textit{File Explorer}, \textit{Gmail}, \textit{Word}, and \textit{Zoom}. These tasks can be found in Table~\ref{tab:tasks}. In total, there were $26$ tasks to complete. The selection of these tasks was based on a closely related work~\cite{uckun2022taming}, and these tasks were comparable in difficulty and captured the diversity in UI interactions for screen reader users.

\subsubsection{\textbf{Procedure}}

The experimenter first explained to the participants that the goal of the study was to collect the variations of natural language commands used to perform the task. The study was set up in a way that made it hard for them to discern that they were interacting with the wizard rather than the system. The experiment was conducted using a \textit{Windows} laptop with a JAWS screen reader. During the study, the users leaned forward as if to speak into the microphone and only turned toward the experimenter when they had questions. Finally, many users expressed interest in getting a copy of the system for testing as soon as possible, which also suggests that they preferred interacting with a functional voice-enabled interface.

Since voice-enabled interaction with the desktop applications was a new concept to all the participants, the experimenter provided a demonstration at the beginning of the experiment by performing a simple task and showing a wide range of voice commands that the participants could issue. After the demo session and collection of demographic data, participants were given $10$ minutes to learn and get familiar with the voice-enabled user interface. After the practice and learning stages, participants completed tasks in $5$ applications. The order of the applications was counterbalanced using the Latin square method~\footnote{\url{http://compneurosci.com/wiki/images/9/98/Latin_square_Method.pdf}}. Each study lasted for nearly $1$ hour, and the participants were given monetary compensation for their time and contribution.

\subsubsection{\textbf{Data collection}}

Audio recordings of the participants were collected during the study. These recordings were transcribed, and the participants' natural language commands for each task were collected. 
\clearpage

% \subsubsection{\textbf{Subjective Feedback}}

% All participants preferred using the voice interface instead of JAWS keyboard shortcuts. Participants $P2$, $P4$, $P5$, $P8$, and $P11$ mentioned that using the voice interface was much faster than trying to complete the tasks with JAWS. To quote participant $P8$, ``It was much faster than typing (using JAWS) ... you don't lose your chain of thought.'' Participants $P2$, $P3$, $P7$, $P9$, $P10$, and $P11$ stated that they loved interacting with computer applications without touching the keyboard. $P9$ said, ``It was less work than typing. No keystrokes, less memorization'' while $P7$ and $P10$ mentioned that the voice interface is necessary for people with motor disabilities or who cannot use their hands. 

\subsection{List of Applications used in the few-shot examples dataset}\label{subap:applications}
\begin{table}[h]
    \begin{tabular}{|c|}
        \hline
        \textbf{Application} \\ \hline
        Excel \\ \hline
        Explorer \\ \hline
        Notepad \\ \hline
        Outlook \\ \hline
        Spotify \\ \hline
        Winword \\ \hline
        Wordpad \\ \hline
        Zoom \\ \hline
        Google Docs (Opened in Chrome) \\ \hline
        Amazon (Opened in Firefox) \\ \hline
        Gmail (Opened in Chrome) \\ \hline
    \end{tabular} \\
    \caption{List of Applications used in the few-shot examples dataset.}
    \label{tab:applications}
\end{table}

\end{document}